\documentclass{article}

\usepackage{arxiv}

\usepackage[utf8]{inputenc} 
\usepackage[T1]{fontenc}    
\usepackage{hyperref}       
\usepackage{url}            
\usepackage{booktabs}       
\usepackage{amsfonts}       
\usepackage{nicefrac}       
\usepackage{microtype}      
\usepackage{lipsum}
\usepackage{graphicx}
\usepackage{enumitem}

\graphicspath{ {./images/} }

\title{psiUnity: A Platform for Multimodal Data-Driven XR}

\author{%
  Akhil Ajikumar, Sahil Mayenkar, Steven Yoo, Sakib Reza, and Mohsen Moghaddam\\[16pt]
  Symbiotic \& Augmented Intelligence Laboratory \\
  Georgia Institute of Technology \\
  Atlanta, GA 30332\\[8pt]
  \texttt{akhil.aji;smayenkar3;steven.yoo;sreza32;mohsen.moghaddam@gatech.edu}
}

\begin{document}

\maketitle

\begin{abstract}
Extended reality (XR) research increasingly relies on the ability to stream and synchronize multimodal data between headsets and immersive applications for data-driven interaction and experimentation. However, developers face a critical gap: the Platform for Situated Intelligence (\textbackslash psi), which excels at deterministic temporal alignment and multimodal data management, has been largely inaccessible to the dominant Unity/MRTK ecosystem used for HoloLens development. We introduce \textbf{psiUnity}, an open-source C\# integration that bridges \textbackslash psi’s .NET libraries with Unity 2022.3 and MRTK3 for HoloLens 2. psiUnity enables bidirectional, real-time streaming of head pose, hand tracking, gaze, IMU, audio, and depth sensor data (AHAT and long-throw) with microsecond-level temporal precision—allowing Unity applications to both consume and produce synchronized multimodal data streams. By embedding \textbackslash psi’s native serialization, logging, and temporal coordination directly within Unity’s architecture, psiUnity extends \textbackslash psi beyond its previous StereoKit limitations and empowers the HRI, HCI, and embodied-AI communities to develop reproducible, data-driven XR interactions and experiments within the familiar Unity environment. The integration is available at \href{https://github.com/sailgt/psiUnity}{https://github.com/sailgt/psiUnity}.
\end{abstract}

\keywords{Extended reality, multimodal data fusion, embodied AI, human–computer interaction, Unity engine}

\section{Introduction}
\label{sec:introduction}

Multimodal research in human–robot interaction (HRI), extended reality (XR), and embodied AI increasingly relies on the ability to capture, synchronize, and process heterogeneous data streams such as video, audio, gaze, hand and body tracking, depth sensing, and robot telemetry in real time. However, ensuring temporal alignment and cross-modal synchronization across multiple sensors and computation nodes remains a major technical challenge. Each sensor typically operates with different frame rates, time stamps, and latency characteristics, making it difficult to establish the unified temporal context required for reliable perception, intent estimation, and adaptive interaction modeling.

To address this challenge, Microsoft introduced the Platform for Situated Intelligence (\textbackslash psi), a modular, open-source framework designed to handle multimodal data acquisition, time synchronization, and visualization at scale \cite{bohus2021platform}. \textbackslash psi provides an event-driven processing pipeline with precise temporal alignment, allowing researchers to compose and visualize complex, time-synchronized data flows using tools such as PsiStudio. Its robust design has made it particularly well-suited for multimodal AI and cognitive systems research.

Despite these advantages, \textbackslash psi’s ecosystem has been primarily centered around desktop visualization and limited XR support through StereoKit, a lightweight mixed-reality framework. While effective for prototyping, StereoKit presented significant barriers for developers and researchers seeking to build rich, interactive XR experiences—particularly those leveraging Unity’s advanced XR toolchains, physics engines, and ecosystem integrations with platforms such as HoloLens, AR Foundation, and the Robot Operating System (ROS).

To bridge this gap, we introduce the \textbf{psiUnity Integration Framework}—a native C\# interface that embeds \textbackslash psi’s multimodal processing capabilities directly within the Unity environment. This integration enables researchers to design, deploy, and evaluate XR applications in Unity while preserving \textbackslash psi’s hallmark feature of precise temporal synchronization across heterogeneous data streams. By leveraging \textbackslash psi’s .NET libraries natively within Unity, psiUnity grants full access to \textbackslash psi’s deterministic logging, replay, and visualization infrastructure—eliminating the need for external bridges or custom serialization layers.

This integration allows seamless development, visualization, and analysis of XR-based multimodal interactions without compromising synchronization, scalability, or reproducibility. In doing so, it opens new avenues for data-driven research in context-aware human–robot collaboration, embodied cognition, and adaptive assistance systems (Fig.~\ref{fig:screenshot}). The key contributions of this work are as follows:

\begin{itemize}[itemsep=0pt, topsep=2pt, parsep=2pt, partopsep=1pt]
    \item An open-source Unity integration layer for \textbackslash psi compatible with MRTK3 and HoloLens~2.
    \item Direct integration of \textbackslash psi's .NET libraries enabling native multimodal data synchronization within Unity.
    \item Native \textbackslash psi serialization and temporal coordination for deterministic logging and replay.
    \item Unity prefabs and example scenes for rapid deployment and extension.
\end{itemize}

\section{Related Work}
\label{sec:relatedwork}

Multimodal, time-synchronized data capture remains a central challenge in extended reality (XR) and human--robot interaction (HRI) research. While frameworks such as \textbf{ROS} \cite{quigley2009ros} offer robust middleware for distributed systems, \textbf{\textbackslash psi} \cite{bohus2021platform} is uniquely designed for deterministic temporal alignment across heterogeneous sensor streams—making it particularly suited for reproducible, data-driven experimentation. Yet, \textbackslash psi's adoption has been limited by a lack of integration within mainstream XR development environments. Most XR research applications are built in the \textbf{Unity Engine} using toolchains such as the \textbf{Mixed Reality Toolkit (MRTK)} \cite{mrtk} and \textbf{OpenXR} \cite{openxr}, especially when accessing \textbf{HoloLens~2 Research Mode} data \cite{ungureanu2020hl2rm}. These Unity-based frameworks, however, lack the native high-precision synchronization, logging, and replay infrastructure that \textbackslash psi provides.

This disconnect has left XR researchers without a unified workflow for developing and studying temporally aligned multimodal interactions. The closest related effort, \textbf{the Situated Interactive Guidance, Monitoring, and Assistance (SIGMA)}  system \cite{bohus2024sigma}, demonstrates a powerful \textbackslash psi-based client--server architecture on the HoloLens~2 using TCP/IP for multimodal data streaming. However, SIGMA functions as a self-contained application rather than a general-purpose Unity integration. In contrast, our work---\textbf{psiUnity}---directly fills this gap by providing an open-source framework that embeds \textbackslash psi's time-synchronized processing pipeline within the Unity/MRTK3 ecosystem, enabling researchers to build reproducible, data-driven XR applications with native access to multimodal synchronization and logging.

\section{System Pipeline}

\begin{figure*}[t]
  \centering
  \includegraphics[width=\textwidth]{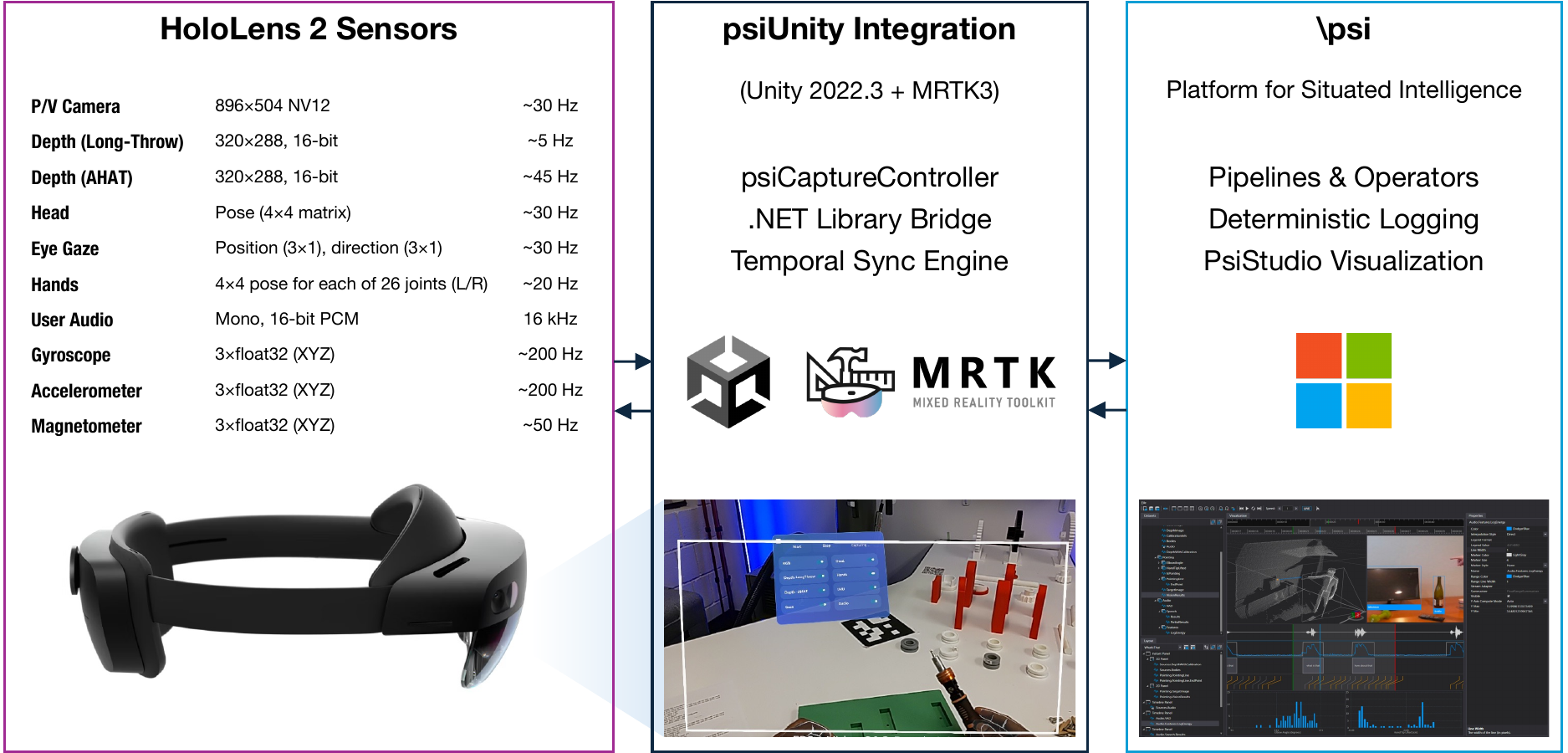}
  \caption{System pipeline and schematic overview of the \textbf{psiUnity} integration framework. 
  The system bridges multimodal sensor data from the HoloLens~2 (Research Mode) into the Unity 2022.3 environment via native C\# components and \textbackslash psi’s .NET libraries.
  The base infrastructure for initializing and managing \textbackslash psi pipelines is provided by the \texttt{PsiCaptureController}, while specific data streams—such as IMU, head pose, eye gaze, hand tracking, RGB video, and depth (AHAT and Long-throw)—are defined and configured within its subclass \texttt{DefaultPsiCaptureController}. The HoloLens 2 captures these diverse sensor streams, which are timestamped and processed directly within Unity using \textbackslash psi’s native serialization format.
  This enables synchronized multimodal logging and replay for XR research, with \textbackslash psi’s deterministic temporal alignment natively embedded into Unity applications. 
  The inset shows a first-person view during the assembly task, where the AR interface allows users to start or stop specific data streams (e.g., RGB, depth, IMU, audio) before initiating multimodal data capture.}
  \label{psiUnity}
\end{figure*}

psiUnity is implemented as a native C\# integration that directly incorporates \textbackslash psi's .NET libraries into the Unity Engine environment (Fig. \ref{psiUnity}). The application is developed in Unity (v. 2022.3) and deployed to the HoloLens 2 as a Universal Windows Platform (UWP) application with an ARM64 architecture. The paramount design goal is the preservation of temporal integrity, ensuring that every data sample is processed by \textbackslash psi with its original, high-precision timestamp captured at the source.

\subsection{Sensor Data Acquisition}

A key enabler for this pipeline is the HoloLens 2 Research Mode, which must be activated on the device. Research Mode provides programmatic access to raw sensor streams that are crucial for many research applications but are not exposed through standard Unity APIs. This includes the dual-mode depth cameras (AHAT and Long-throw) and the Inertial Measurement Unit (IMU). See Fig. \ref{psiUnity}. The feasibility of accessing these streams from a C\# context within Unity has been demonstrated by several open-source projects, which serve as a foundation for this work. Other essential data streams, such as the main color video camera (PV), microphone audio, and high-level spatial inputs like head pose, eye gaze, and articulated hand tracking, are accessed via standard MRTK, Unity, and Windows Media Capture APIs.

\vspace{2em} 

\subsection{\textbackslash psi Integration and Temporal Synchronization}

The cornerstone of the system's scientific validity is its direct integration with \textbackslash psi's temporal coordination engine. As each sensor frame or data sample is captured on the device, it is immediately paired with a high-precision timestamp obtained from the underlying Windows Perception APIs or from standard System timing libraries and passed directly to \textbackslash psi's pipeline components. This ensures that the time of acquisition is recorded with maximum accuracy before any processing latency is introduced. By using \textbackslash psi's native .NET libraries and serialization format, psiUnity leverages \textbackslash psi's proven temporal alignment mechanisms to correctly fuse and synchronize the multiple asynchronous data streams. This native integration eliminates the overhead and complexity of external bridges or custom serialization protocols, providing researchers with \textbackslash psi's full deterministic logging, replay, and visualization capabilities directly within their Unity applications. Although psiUnity integrates natively with \textbackslash psi’s .NET architecture, several modifications were required to underlying \textbackslash psi libraries to ensure compatibility with Unity’s Mono runtime and IL2CPP AOT compilation for UWP.

\section{Conclusion}

In this paper, we introduced \textbf{psiUnity}, an open-source integration framework that connects \textbackslash psi with the Unity Engine for HoloLens~2 applications.  This work addresses a critical gap between \textbackslash psi’s multimodal synchronization capabilities and the widespread adoption of the Unity/MRTK development ecosystem.  By providing a robust pipeline for real-time, time-synchronized streaming of sensor data—including head pose, hand tracking, eye gaze, IMU, audio, and depth—\textbf{psiUnity} enables researchers to leverage \textbackslash psi’s deterministic logging, alignment, and visualization tools directly within modern mixed reality environments.  The framework effectively liberates \textbackslash psi from its previous StereoKit constraints, extending its applicability to a broader community of researchers and developers creating interactive XR experiences in Unity. Ultimately, this integration empowers the HRI, HCI, and embodied-AI communities to design, implement, and evaluate complex multimodal experiments with greater precision, reproducibility, and scalability.

\section{Acknowledgment}
This work was supported by the National Science Foundation under Grant No. 2128743. Any opinions, findings, or conclusions expressed in this material are those of the authors and do not reflect the views of the NSF.

\bibliographystyle{IEEEtran}
\bibliography{references}

@misc{mrtk,
  title        = {Mixed Reality Toolkit (MRTK)},
  howpublished = {\url{https://docs.microsoft.com/en-us/windows/mixed-reality/mrtk-unity}},
  note         = {Accessed: 2025-11-03}
}

@misc{openxr,
  title        = {OpenXR},
  howpublished = {\url{https://www.khronos.org/openxr}},
  note         = {Accessed: 2025-11-03}
}

@article{bohus2021platform,
  title={Platform for situated intelligence},
  author={Bohus, Dan and Andrist, Sean and Feniello, Ashley and Saw, Nick and Jalobeanu, Mihai and Sweeney, Patrick and Thompson, Anne Loomis and Horvitz, Eric},
  journal={arXiv preprint arXiv:2103.15975},
  year={2021}
}

@article{ungureanu2020hl2rm,
  author    = {Ungureanu, Dorin and Bogo, Federica and Galliani, Silvano and Sama, Pooja and Duan, Xin and Meekhof, Casey and Stühmer, Jan and Cashman, Thomas J. and Tekin, Bugra and Schönberger, Johannes L. and Olszta, Pawel and Pollefeys, Marc},
  title     = {HoloLens 2 Research Mode as a Tool for Computer Vision Research},
  journal   = {arXiv preprint},
  year      = {2020},
  volume    = {arXiv:2008.11239}
}

@inproceedings{quigley2009ros,
  author    = {Quigley, Morgan and Conley, Ken and Gerkey, Brian and Faust, Jonathan and Foote, Tully and Leibs, Jeremy and Wheeler, Rob and Ng, Andrew Y.},
  title     = {ROS: an open‐source Robot Operating System},
  booktitle = {ICRA Workshop on Open Source Software},
  volume    = {3},
  pages     = {5},
  year      = {2009},
  address   = {Kobe, Japan}
}

@inproceedings{bohus2024sigma,
  author    = {Bohus, Dan and Andrist, Sean and Saw, Nick and Rad, Mahdi and Paradiso, Ann and Chakraborty, Ishani},
  title     = {{SIGMA}: An Open-Source Interactive System for Mixed-Reality Task Assistance Research - Extended Abstract},
  booktitle = {2024 IEEE Conference on Virtual Reality and 3D User Interfaces Abstracts and Workshops (VRW)},
  pages     = {889--890},
  year      = {2024},
  doi       = {10.1109/VRW62533.2024.00241}
}

\end{document}